\documentclass[aps,pra,superscriptaddress,amsmath,amssymb,twocolumn,showpacs]{revtex4}

\usepackage{graphicx}
\usepackage{color}
\usepackage{bm}

\newcommand{\Ket}[1]{|{#1}\rangle}
\newcommand{\Bra}[1]{\langle{#1}|}
\newcommand{\BraSub}[2]{{}_{#1}\hspace*{-0.2mm}\langle{#2}|}
\newcommand{\BraKet}[2]{\langle{#1}|{#2}\rangle}
\newcommand{\BraKetSub}[3]{{}_{#3}\hspace*{-0.2mm}\langle{#1}|{#2}\rangle_{#3}}
\newcommand{\KetBraSub}[3]{\Ket{#1}_{#3}\hspace*{-0.2mm}\Bra{#2}}

\newcommand{\Sig}[2]{\hat{\sigma}_{#1}^{({#2})}}
\newcommand{\gammae}{\gamma_2}
\newcommand{\gammap}{\gamma_1}
\newcommand{\up}{\uparrow}
\newcommand{\dn}{\downarrow}
\newcommand{\purity}{P}
\newcommand{\totsys}{\text{tot}}
\newcommand{\mstsys}{\text{X}}
\newcommand{\slvsys}{\text{S}}
\newcommand{\rhot}{\rho_\totsys}
\newcommand{\rhos}{\rho} % {\rho^{\slvsys}}
\newcommand{\slave}{\text{S}}
\newcommand{\master}{\text{X}}
\newcommand{\txtslave}{$\slave$}
\newcommand{\txtmaster}{$\master$}
\newcommand{\Vmap}{\mathcal{V}}
\newcommand{\Vtotmap}{\mathcal{V}_{\totsys}}

\newcommand{\tr}{\mathop{\text{tr}}\nolimits}

\newcommand{\sss}[1]{{\bf\em #1} --- }

\newcommand{\todelete}[1]{{\color{red}#1}}

\renewcommand{\todelete}[1]{\, }

\definecolor{dgreen}{rgb}{0,0.5,0}
\definecolor{delete-k}{cmyk}{0.2,0,0,0}

{\, }

\begin{document}
% --------------------  TITLE  --------------------
\title{Influence of dissipation on the extraction of quantum states via repeated measurements}

% ------------  AUTHORS AND AFFILIATIONS ----------
\author{B. Militello}
\email{bdmilite@fisica.unipa.it}
\affiliation{MIUR and Dipartimento di Scienze Fisiche ed Astronomiche dell'Universit\`{a} di Palermo, Via Archirafi 36, I-90123 Palermo, Italy}
\author{K. Yuasa}
\email{kazuya.yuasa@ba.infn.it} \affiliation{Dipartimento di
Fisica, Universit\`a di Bari, I-70126 Bari, Italy}
\affiliation{Istituto Nazionale di Fisica Nucleare, Sezione di
Bari, I-70126 Bari, Italy}
\author{H. Nakazato}
\affiliation{Department of Physics, Waseda University, Tokyo
169-8555, Japan}
\author{A. Messina}
\affiliation{MIUR and Dipartimento di Scienze Fisiche ed Astronomiche dell'Universit\`{a} di Palermo, Via Archirafi 36, I-90123 Palermo, Italy}

%\date{\today}
\date{\today}

% --------------------  ABSTRACT  --------------------

\begin{abstract}
A quantum system put in interaction with another one that is
repeatedly measured is subject to a non-unitary dynamics, through
which it is possible to extract subspaces. This key idea has been
exploited to propose schemes aimed at the generation of pure
quantum states (purification). All such schemes have so far been
considered in the ideal situations of isolated systems. In this
paper, we analyze the influence of non-negligible interactions
with environment during the extraction process, with the scope of
investigating the possibility of purifying the state of a system
in spite of the sources of dissipation. A general framework is
presented and a paradigmatic example consisting of two interacting
spins immersed in a bosonic bath is studied. The effectiveness of
the purification scheme is discussed in terms of purity for
different values of the relevant parameters and in connection with
the bath temperature.
\end{abstract}


\maketitle

% --------------------  INTRODUCTION  ------------------
\section{Introduction}
\label{sec:introduction} During the last decades, enormous
progresses have been made in such physical contexts as cavity
quantum electrodynamics (CQED)~\cite{ref:CQED},
superconductor-based circuits~\cite{ref:SCircuits}, and trapped
ions~\cite{ref:TrapReview}. In connection with the applications in
the field of quantum information~\cite{ref:q_tech}, seminal
experimental results have been reached, such as the
implementations of quantum logical
gates~\cite{ref:NISTCNOT,ref:BlattCNOT,ref:NECCNOT} and the
realizations of quantum teleportation~\cite{ref:Teleportation}.
Generation of quantum states is a crucial issue in
nano-technologies, because it is the basis of initialization
processes with which many experimental protocols begin.
Incidentally, it also provides the possibility of observing
physical systems behaving according to the predictions of quantum
mechanics once a nonclassical state has been created.

Recently, a strategy for the generation of pure quantum states
through extraction from arbitrary initial states has been
proposed~\cite{ref:Nakazato_PRL,ref:Nakazato_PRA}. This procedure
is based on the idea of putting a quantum system in interaction
with another one that is repeatedly measured in order to induce a
non-unitary evolution which forces the former system onto a
Hilbert subspace. If such a subspace is one-dimensional, the
process reduces to the extraction of a pure quantum state. For
this reason, this procedure has been addressed as a
``purification''~\cite{ref:q_tech,ref:Purification}. On the basis
of this general scheme, many applications have been proposed: it
is possible to extract
entanglement~\cite{ref:Nakazato_PRA,ref:Lidar_PRA-Paternostro},
and the initialization of multiple qubits would be useful for
quantum computation~\cite{ref:Nakazato_PRA,ref:qdot}; extensions
of the scheme enable us to establish entanglement between two
spatially separated systems via repeated measurements on an
entanglement mediator~\cite{ref:qpfe-separated}; in single trapped
ions, the extraction of angular-momentum Schr\"odinger-cat states
has been proposed~\cite{ref:Militello_PRA04} and the possibility
of steering the extraction of pure states through quantum Zeno
effect has been predicted~\cite{ref:Militello_PRA05}. In passing,
we mention that the approach here recalled is related to quantum
non-demolition measurements~\cite{ref:general_QND}, introduced for
gravitational wave detection~\cite{ref:GravityWaveDetection} and
exploited in different physical systems for applications in
quantum computation and
information~\cite{ref:QND_QuantumInfoQubits} and for quantum state
generation in general. For instance, in the context of trapped
ions, putting the vibrational degrees of freedom in interaction
with the electronic degrees of freedom, and repeatedly measuring
the atomic state, it is possible to generate Fock states both in
one-dimensional~\cite{ref:TrapOneDim_QND} and two-dimensional
contexts~\cite{ref:TrapTwoDim_QND}.

Until now, the extraction of pure states through repeated
measurements has been considered in ideal situations, that is, the
evolution is assumed to be unitary except for the change of state
by the measurements. In this paper, we discuss how the predictions
change when the system is put in interaction with an environment
and, as a consequence, is subject to a non-unitary evolution (even
between two successive measurements), which is assumed to be
described by a Lindblad-type equation~\cite{ref:MasterEq}. First
of all in this paper, we discuss the behavior of a quantum system,
whose dynamics is governed by the repeated measurements
represented by projection operators on the other system in
interaction with the former and by the dissipative environmental
interaction with the two quantum systems. A criterion for the
extraction of pure states in the presence of dissipation is
derived and a quantum system composed of two two-level systems
(qubits) immersed in a common bosonic bath is analyzed as a very
simple example. The ability or disability of extracting pure
states for a one qubit system by this scheme is estimated
numerically in terms of the purity of the density operator.

The present paper is organized as follows. In
Sec.~\ref{sec:general_sketch}, the general aspects of the
purification protocol in the presence of interaction with an
environment are discussed. In Sec.~\ref{sec:simplest}, a simple
model of two mutually interacting two-level systems immersed in a
bosonic bath is analyzed, and the main differences between the
ideal situation (in the absence of dissipation) and the more
realistic situation here studied (in the presence of dissipation)
are singled out. Finally, some conclusive remarks are given in
Sec.~\ref{sec:conclusion}.

\section{General Framework}\label{sec:general_sketch}

\subsection{Framework}\label{sec:framework-framework}
The scheme of extraction we study in this paper is based on the
idea presented in Ref.~\cite{ref:Nakazato_PRL}. Assume that we are
interested in preparing a target system, \txtslave, into a pure
state. To this end, put it in interaction with an ancilla system,
\txtmaster, which is repeatedly measured and found in the same
state, say $\Ket{\Phi}_\mstsys$. As a consequence of both the
interaction with and the measurements on \txtmaster, system
\txtslave{} is subject to a non-unitary dynamics which forces it
in a subspace. If the subspace is one-dimensional, the relevant
state is extracted. We underline that if a negative result is
obtained measuring system X (i.e., if the ancilla system is found
in a state different from the expected one), then the trial can be
considered as ``failed'' and the system has to be reset in order
to restart the experiment, or we can just continue the process as
a new trial with the initial condition given by the result of the
unsuccessful projection.

In the following, we consider this scheme but in the presence of
dissipation, in order to examine whether a pure state can still be
extracted. To this end, assume now that both \txtslave{} and
\txtmaster{} are immersed in a bath, so that the dynamics of the
whole system S+X between the repeated measurements is governed by
the master equation
\begin{equation}\label{eq:masterequation}
\frac{d\rhot}{dt}=\mathcal{L}\rhot,\qquad
\mathcal{L}=\mathcal{H}+\mathcal{D},
\end{equation}
where $\rhot$ is the density operator of \slave+\master{} (the
trace over the bath degrees of freedom has already been
performed). $\mathcal{H}$ is the Hamiltonian superoperator defined
by $\mathcal{H}\rhot=-i[\hat{H},\rhot]$, while $\mathcal{D}$ is
the Lindblad-type dissipator~\cite{ref:MasterEq} which takes into
account the interaction with the environment, in the Markovian
limit. Let $\mathcal{M}(\tau)=e^{\mathcal{L}\tau}$ denote the
superoperator describing the dissipative dynamics of S+X and
$\mathcal{P}$ the projection superoperator associated with the
measurement of the state $\Ket{\Phi}_\mstsys$, i.e.
\begin{equation}
\mathcal{P}\rhot
=\KetBraSub{\Phi}{\Phi}{\mstsys}\rhot\KetBraSub{\Phi}{\Phi}{\mstsys}.
\end{equation}
Assuming that the time elapsed between two successive measurements
is always $\tau$ and that every of the $N+1$ measurements has
confirmed X to be in the state $\Ket{\Phi}_\text{X}$, the initial
state $\rho_\totsys(0)$ is mapped into the final one,
\begin{equation}\label{eq:nonideal_WOperator}
\rho_\totsys(\tau,N)\propto
\mathcal{P}\mathcal{M}(\tau)\cdots\mathcal{P}\mathcal{M}(\tau)
\mathcal{P}\rho_\totsys(0)=\Vtotmap^N(\tau)\rho_\totsys(0),
\end{equation}
where
\begin{equation}\label{eq:nonideal}
\Vtotmap(\tau)=\mathcal{P}\mathcal{M}(\tau)\mathcal{P},
\end{equation}
which, although rigorously speaking acts on Liouville
space~\cite{ref:Rev_Liouville} of the whole system \slave+\master,
is substantially a superoperator acting on the Liouville space of
subsystem \txtslave, except for the trivial action on \txtmaster{}
(i.e.~projection on the state $\Ket{\Phi}_\mstsys$). Therefore, it
can be rewritten as
$\Vtotmap(\tau)=\Vmap(\tau)\otimes\mathcal{P}$, where
$\Vmap(\tau)$ is a superoperator acting only on the Liouville
space of \txtslave\@.

If the master equation is given in the Lindblad
form~\cite{ref:MasterEq}, the evolution of S+X in interaction with
the bath is written down in the Kraus
representation~\cite{ref:KrausRepr} as
\begin{equation}\label{eq:Kraus}
\mathcal{M}(\tau)\rhot
=\sum_{k=0}^G\hat{T}_k(\tau)\rhot\hat{T}_k^\dag(\tau),
\end{equation}
where $G+1$ is the number of Kraus operators $\hat{T}_k(\tau)$
involved in the decay process. Therefore, the superoperator
$\mathcal{V}(\tau)$ that maps the state of S just after a
measurement to another after the next measurement reads
\begin{equation}\label{eq:KrausContracted}
\Vmap(\tau)\rhos
=\sum_{k=0}^G\hat{V}_k(\tau)\rhos\hat{V}_k^\dag(\tau),
\end{equation}
where $\rho$ is the state of S and
\begin{equation}
\hat{V}_k(\tau)
=\BraSub{\mstsys}{\Phi}\hat{T}_k(\tau)\Ket{\Phi}_\mstsys
\label{eqn:ContracrtedV}
\end{equation}
is an operator acting on the Hilbert space of S\@. Note that the
state of the whole system S+X after a measurement is factorized
like
$\rho_\text{tot}=\rho\otimes\KetBraSub{\Phi}{\Phi}{\text{X}}$.

The properties of $\mathcal{V}(\tau)$ determine the fate of the
state of \txtslave, after many repetitions of measurements on
\txtmaster\@. In fact, assume that $\mathcal{V}(\tau)$ is
diagonalizable and consider its spectral decomposition in terms of
its eigenprojections~\cite{ref:MUIR-Halmos}:
\begin{equation}
\Vmap(\tau)=\sum_n\lambda_n\Pi_n,
\label{eqn:SpectralDecomp}
\end{equation}
where $\Pi_n$ is the eigenprojection belonging to the eigenvalue
$\lambda_n$ and satisfies the ortho-normality and completeness
conditions
\begin{equation}
\Pi_m\Pi_n=\delta_{mn}\Pi_n,\qquad\sum_n\Pi_n=1.
\end{equation}
The eigenvalues $\lambda_n$ are complex-valued in general and
ordered in such a way that
$|\lambda_0|=|\lambda_1|=\cdots=|\lambda_{g_0}|>|\lambda_{g_0+1}|\ge|\lambda_{g_0+2}|\ge\cdots{}$.
If $\mathcal{V}(\tau)$ is not diagonalizable, the Jordan
decomposition applies instead of
(\ref{eqn:SpectralDecomp})~\cite{ref:MUIR-Halmos}. Generalization
of the following argument to such cases is straightforward. See,
for instance, an Appendix of Ref.~\cite{ref:Nakazato_PRA}.

The evaluation of the $N$th power of $\Vmap(\tau)$ shows that, the
larger the number $N$ of the repeated measurements is, the more
dominant the blocks belonging to the maximum (in modulus)
eigenvalues are over the other blocks. That is, for a large enough
$N$, the action of $\Vmap^{N}(\tau)$ diminishes the components of
the system density operator which do not belong to the generalized
eigenspaces corresponding to the maximum (in modulus) eigenvalues.
Indeed,
\begin{equation}\label{eq:VToN_General}
\Vmap^{N}(\tau) =\sum_n\lambda_n^N\Pi_n
\xrightarrow{N\rightarrow\infty}\sum_{n=0}^{g_0} \lambda_n^N\Pi_n,
\end{equation}
which, in the case wherein only one eigenvalue exists whose
modulus is maximum (i.e.~when $g_0=0$), reduces to
\begin{equation}\label{eq:VToN_Special}
\Vmap^{N}(\tau) \xrightarrow{N\rightarrow\infty} \lambda_0^N\Pi_0.
\end{equation}
Therefore, in such a case, for a large enough (depending on the
structure of the spectrum) $N$, the action of $\Vmap^{N}(\tau)$
essentially reduces to the action of $\Pi_0$, hence projecting the
system in the relevant subspace.

As for the outcome state, it is worth mentioning that, if
$\lambda_0$ is degenerated, the state of S extracted by $\Pi_0$
depends on the initial state of \txtslave, $\rhos(0)$, since such
a final state is substantially proportional to $\Pi_0\rhos(0)$,
which depends on $\rhos(0)$ if $\Pi_0$ refers to a
multi-dimensional space. On the contrary, if $\lambda_0$ is not
degenerated and $\Pi_0$ refers to a one-dimensional subspace, i.e.
\begin{equation}
\Pi_0\rhos=f_0(\rhos)\rhos_0
\label{eqn:Pi0}
\end{equation}
with $\rhos_0$ the relevant eigenstate (which is an element of the
Liouville space of \txtslave) and $f_0$ a suitable form on the
Liouville space of \txtslave, the eigenstate $\rhos_0$ is
extracted irrespectively of the initial condition of the system,
provided $\Pi_0\rho(0)\not=0$. As a first step, in this paper we
shall concentrate on such situations wherein there is a single
extracted eigenspace which in addition is one-dimensional.

In an ideal case wherein any sources of dissipation are absent,
Eq.~(\ref{eq:KrausContracted}) reduces to
$\Vmap(\tau)\rhos=\hat{V}_0(\tau)\rhos\hat{V}_0^\dag(\tau)$ with
$\hat{V}_0(\tau)=\BraSub{\mstsys}{\Phi}e^{-i\hat{H}t}\Ket{\Phi}_\mstsys$~\cite{ref:Nakazato_PRL},
so that, if $\hat{V}_0(\tau)$ possesses a nondegenerate and unique
maximum (in modulus) eigenvalue, then the final state to be
extracted is the pure state $\KetBraSub{u_0}{u_0}{\slvsys}$, since
in this case Eq.~(\ref{eqn:Pi0}) is supplemented by
$\rhos_0=\KetBraSub{u_0}{u_0}{\slvsys}$ and
$f_0(\rhos)=\BraSub{\slvsys}{v_0}\rhos\Ket{v_0}_\slvsys$ where
$\Ket{u_0}$ and $\Bra{v_0}$ are the right- and left-eigenvectors
of $\hat{V}_0(\tau)$ belonging to its largest (in modulus)
eigenvalue, respectively. This is the basic idea of the
purification scheme based on the repeated measurements, which was
first proposed in Ref.~\cite{ref:Nakazato_PRL} and has been
analyzed and developed in
Refs.~\cite{ref:Nakazato_PRA,ref:Lidar_PRA-Paternostro,ref:qdot,ref:qpfe-separated,
ref:Militello_PRA04,ref:Militello_PRA05}. On the other hand, it is
important to stress that in the non-ideal case, even if there is a
single and nondegenerate eigenvalue that is maximum in modulus,
this does not guarantee that the state $\rhos_0$ is a pure state
and hence there is no warranty that the final extracted state is
pure, either. In this sense, our state-extraction scheme may not
necessarily be an effective purification scheme. However, we still
try to seek a possibility of extracting a pure state even in the
presence of dissipation. The examination of such situations
wherein we can extract a pure state is the main topic of this
paper.

{\em Efficiency} --- This scheme for extracting quantum states is
a conditional one, in the sense that each time system \txtmaster{}
is measured it has to be found in the same state, denoted by
$\Ket{\Phi}_\mstsys$. In
Refs.~\cite{ref:Nakazato_PRL,ref:Nakazato_PRA,ref:Lidar_PRA-Paternostro,ref:qdot,ref:qpfe-separated,ref:Militello_PRA04,ref:Militello_PRA05}
it is proved that the probability of success of the extraction,
that is the probability of finding system X in the state
$\Ket{\Phi}_\mstsys$ successively $N$ times, is given by the
normalization factor of the state extracted by
(\ref{eq:nonideal_WOperator}),
$\tr_\text{S}\{\Vmap^{N}(\tau)\rhos(0)\}=\sum_n\lambda_n^N
\tr_{\text{S}}\{\Pi_n\rhos(0)\}$, which behaves asymptotically as
$\to\lambda_0^N\tr_\text{S}\{\Pi_0\rhos(0)\}$ as $N\to\infty$ [or
$\to\sum_{n\le g_0}\lambda_n^{N}\tr_\text{S}\{\Pi_n\rhos(0)\}$ in
the more general situation, which from now on we shall not mention
anymore for the sake of simplicity]. These expressions for the
probability of success (still valid in the non-ideal case,
provided the projectors $\Pi_n$ are the appropriate ones) show
that the structure of the spectrum of $\Vmap(\tau)$ plays a
crucial role for the efficiency and fastness of the extraction. In
particular, on the one hand, the fact that $\lambda_0$ has a
modulus quite larger than those of the other eigenvalues makes
$\Vmap^N(\tau)$ quickly approach $\lambda_0^N\Pi_0$, so that a
smaller number of measurements is required to well approximate the
final result $\Pi_0\rhos(0)$. On the other hand, the closer to
unity the modulus of $\lambda_0$ is, the greater the probability
of success is, approaching just $\tr_\text{S}\{\Pi_0\rhos(0)\}$
(without decaying out completely) for $\lambda_0\simeq 1$. On the
contrary, for small values of $\lambda_0$, the probability quickly
approaches zero like $\lambda_0^N$, which means that if a large
number of measurements is required to approach the state
$(\tr_\text{S}\{\Pi_0\rhos(0)\})^{-1}\Pi_0\rhos(0)$ the scheme
becomes very inefficient. Therefore, the number of measurements
necessary to extract the target state would be an important
measure of efficiency. It can be roughly estimated by the
following argument. The idea is to see how much the relevant part
approaching the target state, $\lambda_0^N \Pi_0\rho(0)$,
dominates over the rest:
\begin{equation}\label{eq:DefinitionOfWeight}
p=\frac{\lambda_0^N\|\Pi_0\rho(0)\|}{\sum_n|\lambda_n|^N\|\Pi_n\rho(0)\|}\,,
\end{equation}
where $\|\cdot\|$ is a certain norm, for instance
$\|A\|:={\sqrt{\tr_\text{S}\{A^\dag A\}}}$. This measure
approaches unity in the limit of an infinite number of
measurements, $p\to 1$ as $N\to\infty$. Then, we ask how many
measurements are necessary for this quantity to exceed a desired
value $0<p_0<1$. After rewriting \eqref{eq:DefinitionOfWeight} as
$\lambda_0^N\|\Pi_0\rho(0)\|/(\sum_{n\not=0}|\lambda_n|^N\|\Pi_n\rho(0)\|)=p/(1-p)$,
a sufficient condition for $p\ge p_0$ is given by
\begin{equation}\label{eq:ConditionForN}
\frac{\lambda_0^N\|\Pi_0\rho(0)\|}{(M-1)|\lambda_1|^N R(\rho(0))}
\ge\frac{p_0}{1-p_0},
\end{equation}
where $M$ is the number of eigenvalues or equivalently the
dimension of the Liouville space and
$R(\rho(0))=\max_{n\not=0}\|\Pi_n\rho(0)\|$. The number of
measurements necessary to get a better quality than $p_0$ is
therefore estimated by
\begin{equation}\label{eq:estimateNumbOfMeas}
N\ge \frac{ \ln[p_0/(1-p_0)] +\ln(M-1)
+\ln\frac{R(\rho(0))}{\|\Pi_0\rho(0)\|} }{
\ln|\lambda_0/\lambda_1|}.
\end{equation}
It is important to note that this threshold depends on $\rho(0)$,
according to the expectation that the larger is the norm of the
relevant part in the initial state, $\|\Pi_0\rho(0)\|$, the
smaller is the number of necessary measurements.
%Instead, the dynamics (including the dissipative part) is taken
%into account by the ratio $|\lambda_0/\lambda_1|$ of the two
%biggest eigenvalues of the map.

\subsection{Searching for Pure Eigenvectors: the Criterion}\label{sec:Criterion}
It is easy to show that a necessary and sufficient condition for a
pure state being an eigenvector of the map $\Vmap(\tau)$ is that
it is a simultaneous eigenstate of all the operators
$\hat{V}_k(\tau)$ involved in the relevant map (see
Eq.~(\ref{eq:KrausContracted})). The proof of this statement
proceeds as follows.

$\Longrightarrow$ Obviously, if the state $\Ket{\phi}_\slvsys$ is
a common eigenstate of all $\hat{V}_k$'s, i.e.,
$\hat{V}_k\Ket{\phi}_\slvsys=\alpha_k\Ket{\phi}_\slvsys$ and
consequently
$\BraSub{\slvsys}{\phi}\hat{V}_k^\dag=\BraSub{\slvsys}{\phi}\alpha_k^*$,
then one has
\begin{equation}
\sum_{k=0}^G\hat{V}_k(\tau)
\KetBraSub{\phi}{\phi}{\slvsys}\hat{V}_k^\dag(\tau)
=\lambda_{\phi}\KetBraSub{\phi}{\phi}{\slvsys},
\end{equation}
where
\begin{equation}
\lambda_\phi=\sum_k\alpha_k^*\alpha_k\ge0.
\end{equation}

$\Longleftarrow$ Let the pure state
$\KetBraSub{\phi}{\phi}{\slvsys}$ be an eigenvector of
$\Vmap(\tau)$, then
$\sum_{k=0}^G\hat{V}_k(\tau)\KetBraSub{\phi}{\phi}{\slvsys}
\hat{V}_k^\dag(\tau)=\lambda_\phi\KetBraSub{\phi}{\phi}{\slvsys}$.
Consider now the overlap with a quantum state
$\KetBraSub{\phi_\perp}{\phi_\perp}{\slvsys}$ orthogonal to
$\KetBraSub{\phi}{\phi}{\text{S}}$:
\begin{align}
0&=\lambda_\phi\,\BraKetSub{\phi_\perp}{\phi}{\slvsys}\hspace*{-0.2mm}\BraKet{\phi}{\phi_\perp}_\slvsys\nonumber\\
&=\sum_{k=0}^G\BraSub{\slvsys}{\phi_\perp}\hat{V}_k(\tau)\KetBraSub{\phi}{\phi}{\slvsys}\hat{V}_k^\dag(\tau)\Ket{\phi_\perp}_\slvsys\nonumber\displaybreak[0]\\
&=\sum_{k=0}^G|\BraSub{\slvsys}{\phi_\perp}\hat{V}_k(\tau)\Ket{\phi}_\slvsys|^2,
\end{align}
from which it follows that
$\BraSub{\slvsys}{\phi_\perp}\hat{V}_k(\tau)\Ket{\phi}_\slvsys=0$
for all $k$ and whatever the state $\Ket{\phi_\perp}_\slvsys$ is,
provided it is orthogonal to $\Ket{\phi}_\slvsys$. In other words,
it means that
\begin{equation}
\hat{V}_k(\tau)\Ket{\phi}_\slvsys
=\alpha_k\Ket{\phi}_\slvsys,\quad\forall k,
\label{eq:LookForPure_Chain}
\end{equation}
where $\alpha_k$ is a suitable complex number.
This completes the proof.

\subsection{Searching for Pure Eigenvectors: Purity}
In those cases in which we are not able to extract an exactly pure
state, there is a possibility of extracting ``almost pure''
states, that is, mixed states very close (in the sense of purity)
to the pure states. To look for almost pure states which can be
extracted, let us recall a measure of purity of a given state. We
show later how the purity of the eigenstate of the linear map
$\Vmap(\tau)$ corresponding to the maximum eigenvalue behaves as a
function of the parameters of the scheme, i.e.~the interval of
time $\tau$ and the repeatedly measured state of \txtmaster,
$\Ket{\Phi}_\mstsys$.

The purity of a state is defined as the trace of the square of the
relevant (normalized) density operator~\cite{ref:PurityDef}:
\begin{equation}\label{ref:puritydef}
\purity(\rho)=\tr_\text{S}\rho^2.
\end{equation}
This quantity is upper and lower bounded in accordance with
$1/L\le \purity(\rho)\le1$ with $L$ being the number of levels of
the system under scrutiny. Observe that the maximum value
[$\purity(\rho)=1$] corresponds to pure states, while the minimum
value [$\purity(\rho)=1/L$] corresponds to maximally mixed states
with maximal von Neumann's entropy.

\subsection{Weak-Damping Case}\label{sec:PerturbativeTreatment}
It is possible to derive a formula for the purity of the extracted
state for general systems in the weak-damping regime. Such a
formula would be useful for understanding which parameters spoil
the purity and convenient for an optimization of the purification.

Let us decompose the relevant map $\mathcal{V}(\tau)$ into two
parts,
\begin{equation}
\mathcal{V}(\tau)=\mathcal{V}^{(0)}(\tau)+\delta\mathcal{V}(\tau),
\end{equation}
where $\mathcal{V}^{(0)}(\tau)$ is the map in the absence of the
environmental perturbation and the rest is treated as a
perturbation to it, which is given in the weak-damping regime by
\begin{equation}
\delta\mathcal{V}(\tau)\otimes\mathcal{P} \simeq\int_0^\tau
dt\,\mathcal{P}e^{\mathcal{H}(\tau-t)}
\mathcal{D}e^{\mathcal{H}t}\mathcal{P}.
\end{equation}

Assuming that $\hat{V}_0(\tau)$ is diagonalizable, let
$\Ket{u_n}_\text{S}$ and $\BraSub{\text{S}}{v_n}$ denote its
right- and left-eigenvectors, respectively, which form a complete
ortho-normal set,
$\sum_n\KetBraSub{u_n}{v_n}{\text{S}}=\openone_\text{S}$
\cite{note:degeneracy}. (We also normalize the right-eigenvectors
as $\BraKetSub{u_n}{u_n}{\text{S}}=1$.) Then, the
right-eigenvectors of the ideal map read
\begin{equation}
\mathcal{V}^{(0)}(\tau)\sigma_{mn}^{(0)}
=\lambda_{mn}^{(0)}\sigma_{mn}^{(0)},\quad
\sigma_{mn}^{(0)}=\KetBraSub{u_m}{u_n}{\text{S}}.
\end{equation}
These are orthogonal to the left-eigenvectors
\begin{equation}
\tilde{\sigma}_{mn}^{(0)}=\KetBraSub{v_m}{v_n}{\text{S}}
\end{equation}
in the sense
\begin{equation}
(\tilde{\sigma}_{mn}^{(0)},\sigma_{m'n'}^{(0)})=\delta_{mm'}\delta_{nn'},\quad
(A,B)=\tr_\text{S}\{A^\dag B\}.
\end{equation}
We are interested in a situation where we can purify S in the
absence of the environmental perturbation. That is,
$\lambda_{00}^{(0)}$ is not degenerated and is the only eigenvalue
that is the largest in modulus.

Now, the standard perturbative treatment yields the first-order
correction to the right-eigenvector,
\begin{multline}\label{eqn:DeltaSigma}
\delta\sigma_{mn}^{(1)}=-\sum_{m'n'\neq
mn}\sigma_{m'n'}^{(0)}\frac{(\tilde{\sigma}_{m'n'}^{(0)},\delta\mathcal{V}(\tau)\sigma_{mn}^{(0)})}{\lambda_{m'n'}^{(0)}-\lambda_{mn}^{(0)}}
\,+\,c\,\sigma_{mn}^{(0)}\,,
\end{multline}
where the constant $c$ is set equal to zero by the normalization
condition
$(\tilde{\sigma}_{mn}^{(0)}+\delta\tilde{\sigma}_{mn}^{(1)},\sigma_{mn}^{(0)}+\delta\sigma_{mn}^{(1)})=1$.

This formula is valid when $\lambda_{mn}^{(0)}$ is not
degenerated. We are interested in the state to be extracted,
i.e.~$\rho_0\simeq(\sigma_{00}^{(0)}+\delta\sigma_{00}^{(1)})/(1+\tr_\text{S}\delta\sigma_{00}^{(1)})$.
Since $\lambda_{00}^{(0)}$ has been assumed to be nondegenerated,
the formula (\ref{eqn:DeltaSigma}) is valid for
$\delta\sigma_{00}^{(1)}$. The purity of $\rho_0$ up to this order
is therefore given by
\begin{align}
P(\rho_0) &\simeq1-2\,\bigl( \tr_\text{S}\delta\sigma_{00}^{(1)}
-\BraSub{\text{S}}{u_0}\delta\sigma_{00}^{(1)}\Ket{u_0}_\text{S}
\bigr)\nonumber\displaybreak[0]\\
&=1-2\sum_{mn\neq 00}
\BraSub{\text{S}}{u_n}\hat{Q}_0\Ket{u_m}_\text{S}
\frac{(\tilde{\sigma}_{mn}^{(0)},\delta\mathcal{V}(\tau)\sigma_{00}^{(0)})}{\lambda_{00}^{(0)}-\lambda_{mn}^{(0)}},
\label{eqn:PerturbativePurity}
\end{align}
where $\hat{Q}_0=\openone_\text{S}-\KetBraSub{u_0}{u_0}{\text{S}}$
is a projection operator.

This is the formula for the purity of the extracted state up to
the first order in the decay constants in the weak-damping regime.
This shows that, if the state $\sigma_{00}^{(0)}$ to be extracted
in the ideal case is an eigenstate of the perturbation,
i.e.~$\delta\mathcal{V}(\tau)\sigma_{00}^{(0)}\propto\sigma_{00}^{(0)}$,
the first-order correction to the purity vanishes and the
purification is robust against the environmental perturbation, at
least up to this order. This is a weaker version of the criterion
discussed in Sec.~\ref{sec:Criterion} and is convenient since the
dissipator of a master equation, $\mathcal{D}$, suffices to this
criterion without knowing the Kraus operators $\hat{T}_k(\tau)$ of
the decay process, which may require solving the master equation.
Furthermore, this formula would be useful for finding a parameter
set that optimizes the purity (minimizes the first-order
correction to the purity).

When S is a two-level system, the formula
(\ref{eqn:PerturbativePurity}) is reduced to
\begin{equation}
P(\rho_0) \simeq1-\frac{2}{\BraKetSub{v_1}{v_1}{\text{S}}}
\frac{(\tilde{\sigma}_{11}^{(0)},\delta\mathcal{V}(\tau)\sigma_{00}^{(0)})}{\lambda_{00}^{(0)}-\lambda_{11}^{(0)}}.
\label{eqn:PerturbativePurityTwoLevel}
\end{equation}

\section{A Simple Model}\label{sec:simplest}
In this section, we apply the ideas presented above to the case of
a simple model. Such a system consists of two mutually-interacting
spins immersed in a bosonic bath, one of which is repeatedly
measured to purify the other.

\subsection{Model}
\sss{Two-spin system} Consider a system of two interacting spins
or pseudo-spins, for instance a couple of identical two-level
atoms subjected to a dipolar coupling. Assuming that the matrix
elements of the dipole operators are real, and neglecting the
counter-rotating terms, one reaches the following Hamiltonian (for
details, see
Refs.~\cite{ref:Nakazato_PRA,ref:TwoSpinsInteracting}):
\begin{equation}
\hat{H}_\totsys
=\sum_{i=\text{S},\text{X}}\frac{\Omega}{2}(1+\Sig{z}{i})
+\epsilon(\Sig{+}{\slvsys}\Sig{-}{\mstsys}
+\Sig{-}{\slvsys}\Sig{+}{\mstsys})
\end{equation}
where
$\Sig{z}{i}=\KetBraSub{\up}{\up}{i}-\KetBraSub{\dn}{\dn}{i}$,
$\Sig{+}{i}=\KetBraSub{\up}{\dn}{i}=(\Sig{-}{i})^\dag$, $\Omega$
is the Bohr frequency of the two-level system and $\epsilon$ the
coupling constant. We have set $\hbar=1$.

The eigenstates of the Hamiltonian are the triplet and singlet
two-spin states:
\begin{subequations}
\begin{align}
\Ket{2}_\totsys &=\Ket{\up}_\slvsys\Ket{\up}_\mstsys,\displaybreak[0]\\
\Ket{1}_\totsys &=\frac{1}{\sqrt{2}}\bigl[
\Ket{\up}_\slvsys\Ket{\dn}_\mstsys+\Ket{\dn}_\slvsys\Ket{\up}_\mstsys
\bigr],\\
\Ket{0}_\totsys &=\Ket{\dn}_\slvsys\Ket{\dn}_\mstsys,\\
\Ket{s}_\totsys &=\frac{1}{\sqrt{2}}\bigl[
\Ket{\up}_\slvsys\Ket{\dn}_\mstsys-\Ket{\dn}_\slvsys\Ket{\up}_\mstsys
\bigr],
\end{align}
\end{subequations}
which are common eigenstates of $\hat{\bm{\Sigma}}^2$ and
$\hat{\Sigma}_z$ with
$\hat{\bm{\Sigma}}=\frac{1}{2}(\hat{\bm{\sigma}}^{(\mstsys)}+\hat{\bm{\sigma}}^{(\slvsys)})$,
whose eigenvalues are given by $\Sigma(\Sigma+1)$ and $m_\Sigma$,
respectively. The corresponding eigenenergies are $2\Omega$,
$\Omega+\epsilon$, $0$, and $\Omega-\epsilon$, respectively. If we
consider the case $\Omega>\epsilon$, then $\Ket{0}$ is the ground
state.

\sss{Interaction with a bosonic bath} The interaction with a
bosonic bath, whose free Hamiltonian is given by
$\hat{H}_\text{B}=\int dk\,\omega_k\hat{a}_k^\dag\hat{a}_k$, is
modelled through the system-bath interaction Hamiltonian
\begin{equation}
\hat{H}_\text{I}
=\sum_{i=\slvsys,\mstsys}(\hat{B}_i+\hat{B}_i^\dag)
(\hat{\sigma}_+^{(i)}+\hat{\sigma}_-^{(i)}),
\end{equation}
where $\hat{B}_i=\int dk\,g_k(\bm{r}_i)\hat{a}_k$, with $\bm{r}_i$
the position of spin $i$ and $g_k(\bm{r}_i)$ the coupling constant
between the atom at position $\bm{r}_i$ and bath mode $k$.
Following the standard derivation~\cite{ref:ME_Derivation} and
assuming the spins very close each other in order to have
$g_k(\bm{r}_1)\simeq g_k(\bm{r}_2)$, we reach the following master
equation in the Schr\"odinger picture for the density operator
$\rhot$ of S+X:
\begin{align}
\frac{d\rhot}{dt}
={}&{-i}[\tilde{H}_\totsys,\rhot]\nonumber\\
&{}+\gammae(1+n_-)\mathcal{D}_{21}\rhot
+\gammap(1+n_+)\mathcal{D}_{10}\rhot\nonumber\\
&{}+\gammae n_-\mathcal{D}_{12}\rhot
+\gammap n_+\mathcal{D}_{01}\rhot
\label{eqn:MasterEqModel}
\end{align}
with
$\mathcal{D}_{ij}\rhot=\KetBraSub{j}{i}{\text{tot}}\rhot\KetBraSub{i}{j}{\text{tot}}-\frac{1}{2}\{\KetBraSub{i}{i}{\text{tot}},\rhot\}$,
$n_\pm$ the mean numbers of bosons in the bath modes of
frequencies $\Omega\pm\epsilon$, which are the Bohr frequencies
between the states involved in the transitions
$\Ket{2}_\totsys\to\Ket{1}_\totsys$ ($\Omega-\epsilon$) and
$\Ket{1}_\totsys\to\Ket{0}_\totsys$ ($\Omega+\epsilon$). $\gammap$
and $\gammae$ are the decay rates related to such modes evaluated
as the spectral correlation functions of
$\hat{B}_i+\hat{B}_i^\dag$, and are related to $g_k$'s by
$\gamma_1=2\pi\int\,dk\,|g_k|^2\delta(\omega_k-\Omega+\epsilon)$
and
$\gamma_2=2\pi\int\,dk\,|g_k|^2\delta(\omega_k-\Omega-\epsilon)$.
Finally, $\tilde{H}_\totsys$ is the Lamb-shifted Hamiltonian of
\slave+\master\@.

\subsection{Extraction of Pure States under the Influence of a Zero-Temperature Bosonic Bath}\label{sec:very_special_case}
Consider now the special case wherein the bath is at zero
temperature.  The spin labelled with \txtmaster{} is repeatedly
measured and found in the state
$\Ket{\Phi}_\mstsys=\cos\frac{\theta}{2}\Ket{\up}_\mstsys+e^{i\chi}\sin\frac{\theta}{2}\Ket{\dn}_\mstsys$,
while the other spin, labelled with \txtslave, is driven toward a
quantum state through its interaction with \mstsys\@. The same
situation is discussed in Ref.~\cite{ref:Nakazato_PRA}, in the
absence of the environmental coupling, where it is found that the
extracted state can be made pure very efficiently, in particular
measuring the states $\Ket{\up}_\text{X}$ and
$\Ket{\dn}_\text{X}$. The evolution of the damped system between
two successive measurements is easily evaluated, for instance,
following the approach developed in Ref.~\cite{ref:Nakazato_LANL},
and is given by (see Eq.~\ref{eq:Kraus})
\begin{subequations}
\label{eq:MESolution}
\begin{equation}
\rhot(t)=\sum_{k=0}^3\hat{T}_k(t)\rhot(0)\hat{T}_k^\dag(t)
\end{equation}
with 4 Kraus operators,
\begin{align}
\hat{T}_0(t)
={}&\KetBraSub{0}{0}{\totsys}
+e^{-\frac{\gammap}{2}t}e^{-i(\Omega+\epsilon)t}\KetBraSub{1}{1}{\totsys}
\nonumber\\
&{}+e^{-\frac{\gammae}{2}t}e^{-i2\Omega t}\KetBraSub{2}{2}{\totsys}
+e^{-i(\Omega-\epsilon)t}\KetBraSub{s}{s}{\totsys},
\label{eq:MESolutionAppendixT0}\\
\hat{T}_1(t)
={}&\sqrt{1-e^{-\gammap t}}\KetBraSub{0}{1}{\totsys},
\label{eq:MESolutionAppendixT1}\displaybreak[0]\\
\hat{T}_2(t)
={}&\sqrt{\frac{\gammae}{\gammap-\gammae}
(e^{-\gammae t}-e^{-\gammap t})}\KetBraSub{1}{2}{\totsys},
\label{eq:MESolutionAppendixT2}\displaybreak[0]\\
\hat{T}_3(t)
={}&\sqrt{1+\frac{\gammae e^{-\gammap t}-\gammap e^{-\gammae t}}{\gammap-\gammae}}\KetBraSub{0}{2}{\totsys}.
\label{eq:MESolutionAppendixT3}
\end{align}
\end{subequations}
$\hat{T}_0(t)$ reduces to the unitary evolution operator in the
case $\gammae=\gammap=0$, whereas the others, i.e.~$\hat{T}_k(t)$
for $k\ge 1$, vanish.

Now measure system \txtmaster{} repeatedly every after $\tau$
during the dissipative dynamics (\ref{eq:MESolution}). According
to (\ref{eq:LookForPure_Chain}), in order for the \txtslave{}
state $\KetBraSub{\phi}{\phi}{\slvsys}$, with
$\Ket{\phi}_\slvsys=\cos\frac{\eta}{2}\Ket{\up}_\slvsys+e^{i\xi}\sin\frac{\eta}{2}\Ket{\dn}_\slvsys$,
be a pure eigenstate of the contracted map $\mathcal{V}(\tau)$, it
should satisfy
\begin{equation}
s_k(\tau)\equiv\BraSub{\slvsys}{\phi_\perp}\BraSub{\mstsys}{\Phi}\hat{T}_k(t)\Ket{\Phi}_\mstsys\Ket{\phi}_\slvsys=0,\quad k=0,1,2,3
\end{equation}
with
$\Ket{\phi_\perp}_\slvsys=\sin\frac{\eta}{2}\Ket{\up}_\slvsys-e^{i\xi}\cos\frac{\eta}{2}\Ket{\dn}_\slvsys$.
Indeed, it is equivalent to look for the eigenstates of the
contracted operators $\hat{V}_k(\tau)$ defined in
(\ref{eqn:ContracrtedV}). It is straightforward to find that
\begin{equation}
\begin{cases}
\medskip
s_1(\tau)
\propto\sin\frac{\theta}{2}\cos\frac{\eta}{2}\left(
e^{i\xi}\cos\frac{\theta}{2}\sin\frac{\eta}{2}
+e^{i\chi}\sin\frac{\theta}{2}\cos\frac{\eta}{2}
\right),\\
\medskip
s_2(\tau)
\propto\cos\frac{\theta}{2}\cos\frac{\eta}{2}\left(
e^{-i\chi}\sin\frac{\theta}{2}\sin\frac{\eta}{2}
-e^{-i\xi}\cos\frac{\theta}{2}\cos\frac{\eta}{2}
\right),\\
s_3(\tau)
\propto\sin\frac{\theta}{2}\cos\frac{\theta}{2}
\sin\frac{\eta}{2}\cos\frac{\eta}{2},
\end{cases}
\end{equation}
where the proportionality factors are the nonvanishing
coefficients in
(\ref{eq:MESolutionAppendixT1})--(\ref{eq:MESolutionAppendixT3}).
From these expressions, it follows that
$s_1(\tau)=s_2(\tau)=s_3(\tau)=0$ is accomplished only for
$\cos\frac{\eta}{2}=0$. This condition is necessary and sufficient
to make $s_k$'s vanish for $k=1,2,3$. In order to make $s_0(t)$
vanish too, it is necessary to have $\sin\frac{\theta}{2}=0$ or
$\cos\frac{\theta}{2}=0$. In fact, the condition
$\cos\frac{\eta}{2}=0$ means
$\Ket{\phi}_\slvsys=\Ket{\dn}_\slvsys$, and evaluating $s_0(t)$ in
such a special situation provides
%\begin{widetext}
\begin{align}
s_0(\tau)
&=\BraSub{\slvsys}{\up}\BraSub{\mstsys}{\Phi}\hat{T}_0(\tau)\Ket{\Phi}_\mstsys\Ket{\dn}_\slvsys\nonumber\displaybreak[0]\\
&=-\frac{e^{-i\chi}e^{-i(\Omega-\epsilon)\tau}}{\sqrt{2}}\sin\frac{\theta}{2}\cos\frac{\theta}{2}
(1-e^{-\frac{\gammap}{2}\tau}e^{-i2\epsilon\tau}),
\end{align}
%\end{widetext}
which, for $\tau\not=0$, vanishes only if $\sin\frac{\theta}{2}=0$ or $\cos\frac{\theta}{2}=0$.

This analysis shows that in some special cases, that is, when the
state of X is repeatedly measured and found in
$\Ket{\Phi}_\mstsys=\Ket{\up}_\mstsys$ ($\sin\frac{\theta}{2}=0$)
or $\Ket{\Phi}_\mstsys=\Ket{\dn}_\mstsys$
($\cos\frac{\theta}{2}=0$), the contracted linear map
$\mathcal{V}(\tau)$ has the \txtslave{} state
$\KetBraSub{\dn}{\dn}{\slvsys}$ as a pure eigenstate. To reach the
final conclusion about the possibility of extracting such a pure
state, one needs to know whether the corresponding eigenvalue is
the maximum (in modulus) in the spectrum of the map.  We shall
therefore diagonalize the contracted map in the two cases,
$\sin\frac{\theta}{2}=0$ and $\cos\frac{\theta}{2}=0$.

Representing the density operator of \txtslave{} as a
four-dimensional vector,
$\rhos=(\rho_{\up\up},\rho_{\dn\dn},\rho_{\up\dn},\rho_{\dn\up})=(\BraSub{\slvsys}{\up}\rhos\Ket{\up}_\slvsys,\BraSub{\slvsys}{\dn}\rhos\Ket{\dn}_\slvsys,\BraSub{\slvsys}{\up}\rhos\Ket{\dn}_\slvsys,\BraSub{\slvsys}{\dn}\rhos\Ket{\up}_\slvsys)$,
the contracted linear map $\Vmap(\tau)$ is substantially
represented by a $4\times 4$ matrix.

\sss{Repeatedly measuring $\Ket{\downarrow}_\text{X}$
($\theta=\pi$)} In the case where the \txtmaster{} state
$\Ket{\dn}_\mstsys$ is repeatedly measured, the corresponding
linear map $\Vmap(\tau)=\Vmap_\dn(\tau)$ is represented by the
following matrix:
\begin{equation}\label{eq:VTauDown}
\Vmap_\dn(\tau)=
\begin{pmatrix}
|f_\dn(\tau)|^2 & 0 & 0 & 0 \\
%& & & \\
\frac{1}{2}(1-e^{-\gammap \tau}) & 1 & 0 & 0 \\
%& & & \\
0 & 0 & f_\dn(\tau) & 0 \\
%& & & \\
0 & 0 & 0 & f_\dn^*(\tau)
\end{pmatrix},
\end{equation}
with
$f_\dn(\tau)=\frac{1}{2}e^{-i(\Omega-\epsilon)\tau}(1+e^{-\frac{\gammap}{2}\tau}e^{-i2\epsilon\tau})$.
The eigenvalues of this matrix are
\begin{equation}
\lambda_0=1,\quad
\lambda_1=\lambda_2^*=f_\dn(\tau),\quad
%\lambda_2=f_\dn^*(\tau),\quad
\lambda_3=|f_\dn(\tau)|^2.
\end{equation}
The right-eigenvector corresponding to the maximum eigenvalue
$\lambda_{0}=1$ is the pure state
$\rho_{0}=\KetBraSub{\dn}{\dn}{\slvsys}$. The larger the time
$\tau$ is, the smaller the other three eigenvalues of the map are
and the faster the extraction of
$\rho_{0}=\KetBraSub{\downarrow}{\downarrow}{\text{S}}$ is, in the
sense that it requires a smaller number of steps.

\sss{Repeatedly measuring $\Ket{\uparrow}_\text{X}$ ($\theta=0$)}
In the case wherein the \txtmaster{} state $\Ket{\up}_\mstsys$ is
repeatedly measured, the map $\Vmap(\tau)$ reduces to
\begin{widetext}
\begin{equation} %\label{eq:VTauUp}
\Vmap_\up(\tau)=
\begin{pmatrix}
e^{-\gammae\tau} & 0 & 0 & 0 \\
%& & & \\
\frac{\gammae(e^{-\gammae\tau}-e^{-\gammap\tau})}{2(\gammap-\gammae)}&
|f_\up(\tau)|^2 & 0 & 0 \\
%& & & \\
0 & 0 & e^{-\frac{\gammae}{2}\tau}f_\up(\tau) & 0 \\
%& & & \\
0 & 0 & 0 & e^{-\frac{\gammae}{2}\tau}f_\up^*(\tau)
\end{pmatrix}%
\end{equation}
\end{widetext}
with
$f_\up(\tau)=\frac{1}{2}e^{-i(\Omega+\epsilon)\tau}(1+e^{-\frac{\gammap}{2}\tau}e^{i2\epsilon\tau})$.
This matrix is easily and exactly diagonalized as long as
$e^{-\gamma_2\tau}\neq|f_\uparrow(\tau)|^2$.
There are two cases
in the ordering of its eigenvalues. Case I: if
$e^{-\frac{\gamma_2}{2}\tau}<|f_\uparrow(\tau)|$,
\begin{equation}
\lambda_0=|f_\uparrow(\tau)|^2,\quad
\lambda_1=\lambda_2^*=e^{-\frac{\gamma_2}{2}\tau}f_\uparrow(\tau),\quad
%=e^{-\frac{\gamma_2}{2}\tau}f_\uparrow^*(\tau),\quad
\lambda_3=e^{-\gamma_2\tau}.
\end{equation}
Case II: if $e^{-\frac{\gamma_2}{2}\tau}>|f_\uparrow(\tau)|$,
\begin{equation}
\lambda_0=e^{-\gamma_2\tau},\quad
\lambda_1=\lambda_2^*=e^{-\frac{\gamma_2}{2}\tau}f_\uparrow(\tau),\quad
%=e^{-\frac{\gamma_2}{2}\tau}f_\uparrow^*(\tau),\quad
\lambda_3=|f_\uparrow(\tau)|^2.
\end{equation}
In case I (which surely occurs in the strong-damping regime
$\gamma_2\tau\rightarrow\infty$), a pure state
$\rho^\text{(I)}=\KetBraSub{\dn}{\dn}{\slvsys}$ is extracted,
while in case II, a mixed state is extracted,
\begin{subequations}
\label{eq:SpecialEigenState}
\begin{equation}
\rho^{\text{(II)}} =p_{\up\up}\KetBraSub{\up}{\up}{\slvsys}
+p_{\dn\dn}\KetBraSub{\dn}{\dn}{\slvsys}
\end{equation}
with
\begin{gather}
p_{\up\up}=\frac{1}{1+\alpha},\qquad
p_{\dn\dn}=1-p_{\up\up}=\frac{\alpha}{1+\alpha},\\
\alpha=\frac{\gammae(e^{-\gammae\tau}-e^{-\gammap\tau})}{2(\gammap-\gammae)(e^{-\gammae\tau}-|f_\uparrow(\tau)|^2)}.
\end{gather}
\end{subequations}
The latter is not in contradiction with the previous statement
that one has a pure eigenstate for $\sin\frac{\theta}{2}=0$.
Indeed, the state $\KetBraSub{\dn}{\dn}{\slvsys}$ is still an
eigenstate of the map, but it does not correspond to the maximum
eigenvalue anymore, and then it is not the state to be extracted.

The purity of the state $\rho^{\text{(II)}}$ in
(\ref{eq:SpecialEigenState}) is given by
\begin{equation}
P(\rho^{\text{(II)}}) =p_{\up\up}^2+p_{\dn\dn}^2
=\frac{1+\alpha^2}{(1+\alpha)^2}.
\end{equation}
In the weak-damping case $\gamma_1\tau,\gamma_2\tau\ll1$ (and
assuming $\sin\epsilon\tau\not=0$ for simplicity), one has
$\alpha\simeq\frac{\gammae\tau}{2\sin^2\!\epsilon\tau}$ up to the
first order in $\gamma_1\tau$ and $\gamma_2\tau$, and hence
\begin{equation}
P(\rho^{\text{(II)}})\simeq 1-2\alpha
\simeq1-\frac{\gammae\tau}{\sin^2\!\epsilon\tau}.
\label{eqn:PurityModelWeakDamping}
\end{equation}
This formula, that alternatively can be directly derived using
\eqref{eqn:PerturbativePurityTwoLevel}, shows that in the
weak-damping regime (i) the purity is linearly affected by
$\gamma_2$, while (ii) it is not influenced by $\gamma_1$.
Furthermore, (iii) the purity is optimized by taking a nontrivial
time interval $\tau\simeq0.37\pi/\epsilon$.

It is worth noting that in the weak damping limit we cannot
extract a pure state, while in the strong damping limit a pure
state can be obtained, which is the opposite one would expect. To
understand this fact, consider first of all that
\txtslave$+$\txtmaster\ has two stable states,
$\Ket{0}_\totsys\Bra{0}$ and $\Ket{s}_\totsys\Bra{s}$ according to
\eqref{eq:MESolution}, and second that in the strongly dissipative
case the system has time to relax onto the equilibrium state which
is a mixture of the two stable states, whose statistical weights
are determined by the initial condition. Then, repeatedly
measuring the state $\Ket{\up}_\master$ cuts the population of
$\Ket{0}_\totsys$ in the mixture and leaves only
$_\master\BraKet{\up}{s}_\totsys\BraKet{s}{\up}_\master\propto\KetBraSub{\dn}{\dn}{\slave}$.

The case $e^{-\frac{\gamma_2}{2}\tau}=|f_\up(\tau)|$ corresponds
to a degenerate case and hence, as clarified in
Sec.~\ref{sec:general_sketch}, is not in the scope of this paper
since it does not permit the extraction of a precise state
irrespectively of the initial state of the system.

Notice that the general case corresponding to measuring a generic
state $\Ket{\Phi}_\master$ can be discussed in the weak damping
limit through the perturbation analysis.

\subsection{At Finite Temperature}\label{sec:behavior_purity}
The analysis on the model has so far been focused on the
zero-temperature case and showed that the only pure state that can
be extracted at the zero temperature is $\Ket{\dn}_\slvsys$ when
the state of X is repeatedly measured and found in
$\Ket{\up}_\mstsys$ or $\Ket{\dn}_\mstsys$. The question of what
happens in the case of non-zero temperature naturally arises. To
answer this question, we resort to numerical calculations.

\begin{figure*}
\begin{tabular}{c@{\qquad\qquad}c}
\includegraphics[width=0.45\textwidth, angle=0]{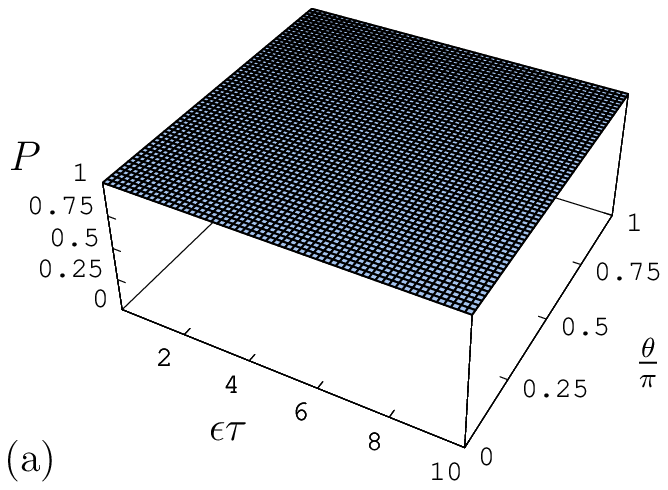}&
\includegraphics[width=0.45\textwidth, angle=0]{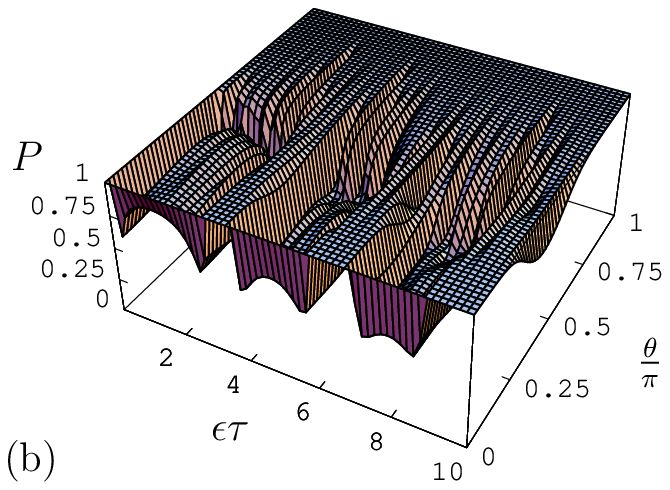}\\
\includegraphics[width=0.45\textwidth, angle=0]{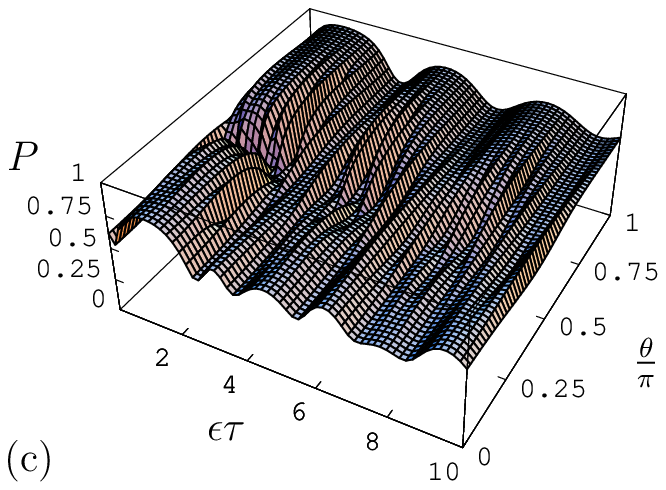}&
\includegraphics[width=0.45\textwidth, angle=0]{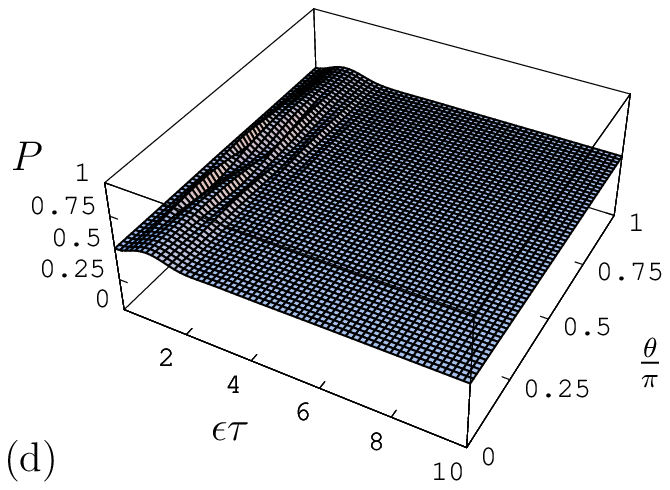}
\end{tabular}
\caption{The purity of the extracted state vs the parameters
$\epsilon\tau$ and $\theta/\pi$ ($\chi=0$ for all cases), in
different situations: (a) in the ideal case, i.e.\ in the absence
of interaction with the bath, and in the presence of interaction
with the bath with (b) ${k_{\text{B}}T}/{\hbar\Omega}=0.01$, (c)
${k_{\text{B}}T}/{\hbar\Omega}=1$, and (d)
${k_{\text{B}}T}/{\hbar\Omega}=10$. In all cases, ratios between
salient physical quantities are fixed as ${\Omega}/{\epsilon}=10$,
${\gammae}/{\epsilon}=0.1$, ${\gammap}/{\gammae}=0.95$.}
\label{fig:Purity_Parameters}
\end{figure*}
The linear map ${\Vmap}(\tau)$ depends on $\tau$, the measured
state $\Ket{\Phi}_{\text{X}}$ (which is individualized in the
Bloch-sphere by the polar and azimuthal angles, $\theta$ and
$\chi$, respectively), and in general the temperature of the
environment, $T$. Given the map, the eigenvector associated with
the maximum eigenvalue, and its purity are functions of all such
quantities ($\tau$, $\theta$, $\chi$ and $T$).

\begin{figure}
\includegraphics[width=0.45\textwidth, angle=0]{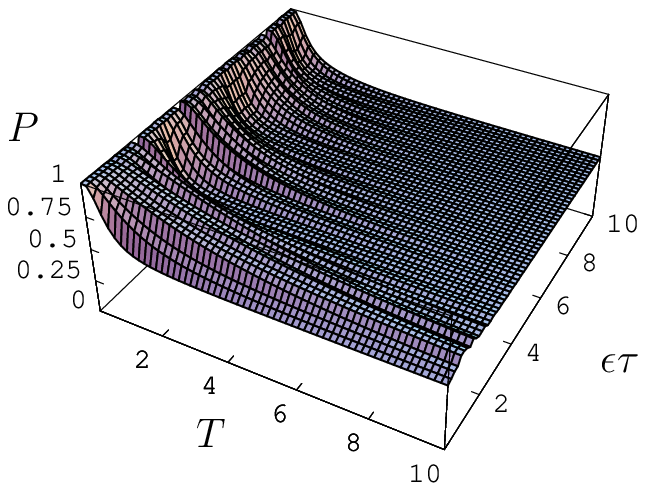}
\caption{The purity of the extracted state as a function of the
temperature $T$ (in units of $\hbar\Omega/k_{\text{B}}$) and
$\epsilon\tau$. The other parameters are: $\theta=3\pi/4$,
$\chi=0$, ${\Omega}/{\epsilon}=10$, ${\gammae}/{\epsilon}=0.1$,
${\gammap}/{\gammae}=0.95$.} \label{fig:Purity_vs_Temperature}
\end{figure}

In Fig.~\ref{fig:Purity_Parameters}, the purity of the state to be
extracted is shown as a function of the parameters $\tau$ and
$\theta$ with $\chi=0$ being fixed, in four different situations
concerning the bath. For the numerical calculations, we have set
${\Omega}/{\epsilon}=10$, ${\gammae}/{\epsilon}=0.1$, and
${\gammap}/{\gammae}=0.95$.

In Fig.~\ref{fig:Purity_Parameters}(a), the purity of the state to
be extracted in the ideal situation is plotted, that is, in the
absence of interaction with the bath. According to the discussion
in Sec.~\ref{sec:general_sketch}, the purity in such a case is
expected to be equal to $1$ whatever the parameters $\tau$,
$\theta$ and $\chi$ are. In the other three figures, the behavior
of purity in the presence of interaction with the bath at
different temperatures is shown.
Figure~\ref{fig:Purity_Parameters}(b) refers to the case of very
low temperature, effectively zero, and shows that in some regions
of the parameter space, there is a possibility of extracting pure
or almost pure states. This is not in contradiction with the
analysis in Sec.~\ref{sec:very_special_case}, where it has been
found that a pure state can be extracted only for $\theta=0,\pi$.
This result refers to an exactly pure eigenstate, while the
numerical calculations here reported show the value of purity,
which can be very close to unity although not exactly $1$.

In Fig.~\ref{fig:Purity_Parameters}(c), one can see that in
practice there is no region in the parameter space corresponding
to pure states: the purity is visibly smaller than unity
everywhere. Finally, in Fig.~\ref{fig:Purity_Parameters}(d), we
see that at a higher temperature ($k_{\text{B}}T=10\hbar\Omega$,
with $k_{\text{B}}$ the Boltzmann constant), the purity of the
state to be extracted is equal to the minimum value for the
two-level system, $\frac{1}{2}$, almost everywhere, that is,
irrespectively of the values of parameters $\tau$ and $\theta$.

In Fig.~\ref{fig:Purity_vs_Temperature}, the purity is plotted as
a function of the temperature $T$ and of the time interval $\tau$
between successive measurements, when a fixed state of system
\txtmaster{} characterized by $\theta=3\pi/4$ and $\chi=0$ is
repeatedly measured. It is well visible that the more the
temperature increases, the more the purity of the extracted state
approaches the minimum value, that is, $\frac{1}{2}$.

All these results express in a clear way that the interaction with
an environment deteriorates the reliability of the purification
scheme based on repeated measurements, although at the zero
temperature pure states can still be extracted.

\subsection{Efficiency}
In a realistic situation, the probability of extracting the target
state as well as the number of measurements one has to perform are
important factors to consider. According to the discussion at the
end of Sec.~\ref{sec:framework-framework}, the probability of
success is asymptotically given by
$\lambda_0^N\tr_\text{S}\{\Pi_0\rhos(0)\}$. Therefore, except for
those situations wherein the maximum eigenvalue (in modulus) is
unity, the most relevant condition to get a good efficiency is
that the number of required measurements is very low, which
implies $|\lambda_1/\lambda_0|\ll 1$, or, better, that the
denominator in the threshold given in
\eqref{eq:estimateNumbOfMeas} is high,
i.e.~$\ln\left|\lambda_0/\lambda_1\right|\gg 1$. Therefore, the
peaks of $\ln\left|\lambda_0/\lambda_1\right|$ correspond to the
maxima of the efficiency (i.e., the minima of the required number
of measurements). To better fix the idea, if we ask that the
target state is obtained with a precision $p_0=0.99$, since we
have $\ln(4-1)+\ln[p_0/(1-p_0)]\approx 5.7$ we find that, in
correspondence to those peaks wherein
$\ln\left|\lambda_0/\lambda_1\right|\sim 4$, the process requires
one or two measurements when the system starts with an initial
condition satisfying $\|\Pi_0\rho(0)\|\sim R(\rho(0))$, which for
instance is usually the case for the maximally mixed state.

\begin{figure*}
\begin{tabular}{c@{\qquad\qquad}c}
\includegraphics[width=0.45\textwidth, angle=0]{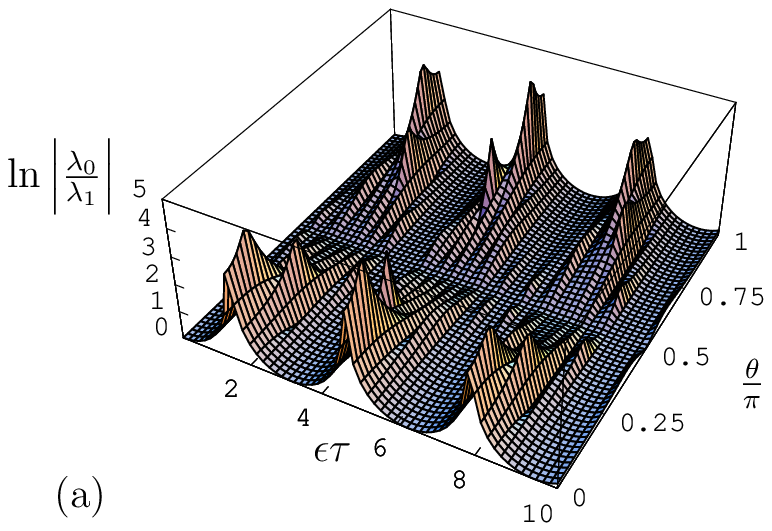}&
\includegraphics[width=0.45\textwidth, angle=0]{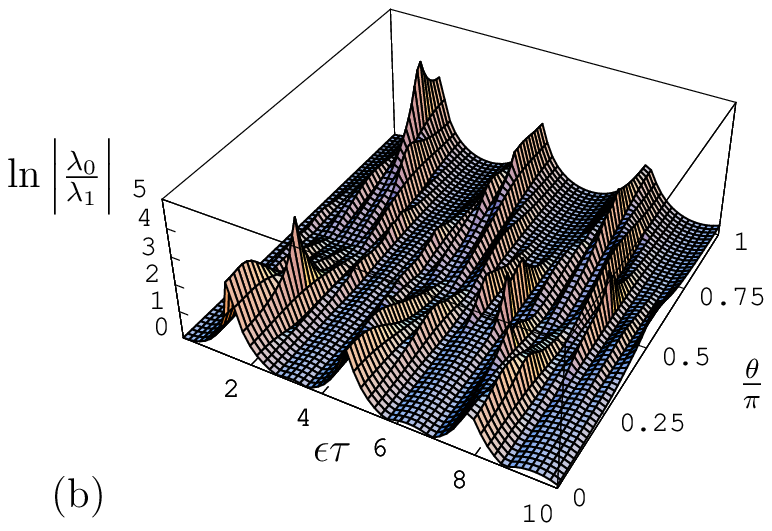}\\
\includegraphics[width=0.45\textwidth, angle=0]{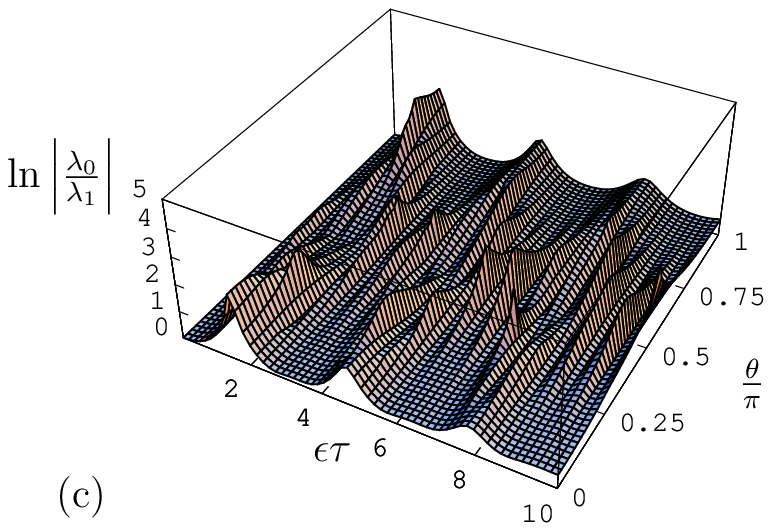}&
\includegraphics[width=0.45\textwidth, angle=0]{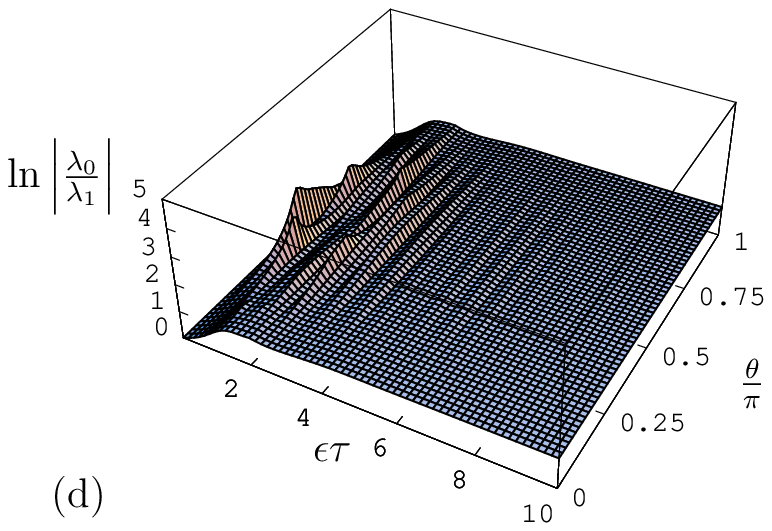}
\end{tabular}
\caption{The quantity $\ln\left|\lambda_0/\lambda_1\right|$ vs the
parameters $\epsilon\tau$ and $\theta/\pi$ ($\chi=0$ for all
cases), in different situations: (a) in the ideal case, i.e.\ in
the absence of interaction with the bath, and in the presence of
interaction with the bath with (b)
${k_{\text{B}}T}/{\hbar\Omega}=0.01$, (c)
${k_{\text{B}}T}/{\hbar\Omega}=1$, and (d)
${k_{\text{B}}T}/{\hbar\Omega}=10$. In all cases, ratios between
salient physical quantities are fixed as ${\Omega}/{\epsilon}=10$,
${\gammae}/{\epsilon}=0.1$, ${\gammap}/{\gammae}=0.95$.}
\label{fig:Efficiency}
\end{figure*}

In Fig.~\ref{fig:Efficiency}(a), we consider the ideal case, while
in Figs.~\ref{fig:Efficiency}(b)--\ref{fig:Efficiency}(d), we
refer to non-ideal situations at zero, intermediate, and high
temperature. The plots clearly show that the interaction with a
nonzero temperature environment negatively affects the efficiency,
lowering the peaks and extending the valleys. Nevertheless, at
zero temperature, various peaks are still present, and in fact, at
zero temperature, the degradation with respect to the ideal case is
not so dramatic.

% ----------------  CONCLUSION  -----------------

\section{Summary}\label{sec:conclusion}

Let us summarize the results reported in this paper. Putting a
system in interaction with a repeatedly measured one forces the
former system onto a subspace, hence realizing, under suitable
conditions, the extraction of pure states. In a more realistic
situation, the two systems are interacting with their environment
too, and therefore are subjected to dissipation. Such an
interaction practically reduces the chance to extract pure states.

From the mathematical point of view, the main difference between
the two situations is represented by the fact that in the ideal
case one extracts eigenvectors of a map onto a Hilbert space,
whereas in the non-ideal case one extracts eigenvectors of a map
onto a Liouville space. We have explored the general framework and
studied a very simple physical system (two spins interacting with
a bosonic bath) in order to bring to light fundamental features of
repeated-measurement based extraction processes in the presence of
dissipation. In Sec.~\ref{sec:very_special_case}, we have shown
that a mixed state is extracted instead of a pure state. Actually,
this is what generally happens, especially at high temperatures.
Nevertheless, with a zero-temperature bath, it is still possible
to extract pure and almost pure states (see
Fig.~\ref{fig:Purity_Parameters}(b)) with still fairly good
efficiency.
\begin{figure}
\includegraphics[width=0.48\textwidth, height=0.28\textwidth]{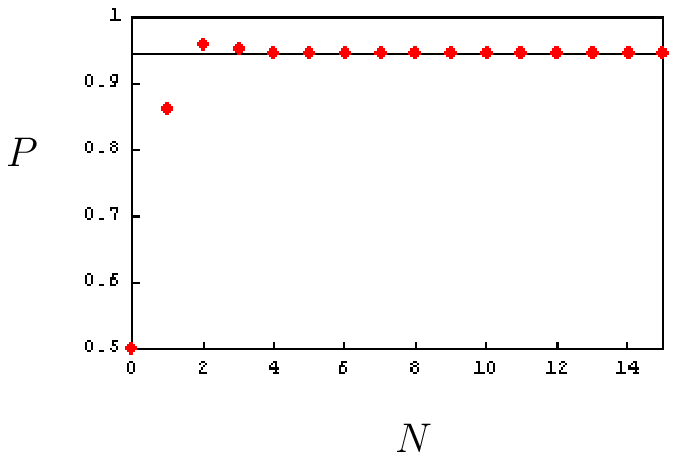}
\caption{The purity of the state of system S as a function of the
number of measurements on X. The initial state is
$\rhos(0)=0.5\KetBraSub{\up}{\up}{\slvsys}+0.5\KetBraSub{\dn}{\dn}{\slvsys}$
and its purity is $0.5$. Observe that the purity reaches the
maximum value at $N=2$ and then decays to an asymptotic value.
Here, the parameters are $\chi=0$, $\theta=2.25$, and
$\epsilon\tau=7.82$, and we have set
$k_{\text{B}}T/\hbar\Omega=10^{-3}$, ${\Omega}/{\epsilon}=10$,
${\gamma_2}/{\epsilon}=0.1$, ${\gamma_1}/{\gamma_2}=0.95$.}
\label{fig:purity_vs_N}
\end{figure}
Indeed, the efficiency, though negatively affected by the
environment, is still good at zero temperature.

Overall, we have considered the case wherein a very large number
of measurements (evaluating the mathematical limit for an infinite
number of measurements) is performed on the ancilla system, as
clearly expressed by (\ref{eq:VToN_General}) and
(\ref{eq:VToN_Special}). We conclude this paper expecting that in
some cases a reduction of the number of measurements performed on
the ancilla system entails an increase of the purity of the output
state. See, for example, Fig.~\ref{fig:purity_vs_N}, wherein we
have plotted the purity of the resulting quantum state as a
function of the number of measurements performed on the ancilla
system, starting from the maximally mixed initial state, with a
particular parameter set. It is well visible that the purity,
starting from its minimum value ($\frac{1}{2}$), increases at the
second measurement, and then decreases down to its asymptotic
value. Therefore, in such a case, the highest value of purity is
obtained for a smaller number of measurements ($N=2$). We will
discuss this aspect of our scheme in the next future.

\section*{ACKNOWLEDGMENTS}
This work is partly supported by the bilateral Italian-Japanese
Projects II04C1AF4E on ``Quantum Information, Computation and
Communication'' of the Italian Ministry of Education, University
and Research, and 15C1 on ``Quantum Information and Computation''
of the Italian Ministry for Foreign Affairs, by the Grant for The
21st Century COE Program at Waseda University and the
Grant-in-Aid % for Scientific Research on Priority Areas (No.\ 13135221) and
for Young Scientists (B) (No.\ 18740250) from the Ministry of
Education, Culture, Sports, Science and Technology, Japan, and by
the Grants-in-Aid for JSPS Postdoctoral Fellowship for Foreign
Researchers (Short-term) and for Scientific Research (C) (No.\
18540292) from the Japan Society for the Promotion of Science. One
of the authors (K.Y.) is supported by the European Union through
the Integrated Project EuroSQIP. Moreover, the authors acknowledge
partial support from University of Palermo in the context of the
bilateral agreement between University of Palermo and Waseda
University, dated May 10, 2004.


\begin{thebibliography}{10}
\bibitem{note:degeneracy}
In this subsection, degenerate eigenvalues, if any, are ``unfolded,'' e.g., a
  three-fold degenerated eigenvalue is given three different subscripts to
  distinguish the three linearly independent vectors belonging to it.
\end{thebibliography}
\end{document}